\documentclass[12pt]{article}
\newcommand{\sect}[1]{\setcounter{equation}{0}\section{#1}}

\evensidemargin=\oddsidemargin
\newcommand{\be}{\begin{equation}}
\newcommand{\ee}{\end{equation}}
\newcommand{\bea}{\begin{eqnarray}}
\newcommand{\eea}{\end{eqnarray}}
\newcommand{\p}{\partial}
\newcommand{\nn}{\nonumber}

\begin{document}
\renewcommand{\thefootnote}{\fnsymbol{footnote}}
\begin{titlepage}
\begin{flushright}
IP/BBSR/2003-42\\
ROM2F/2003/35\\
USITP-03-13\\
hep-th/0312224\\
\end{flushright}
\vskip .5in
\begin{center}
{\large \bf D-branes in the NS5 Near-horizon pp-Wave Background}
\vskip .5in
{\bf S. F. Hassan}$^a$\footnote{e-mail: {\tt fawad@physto.se}},
{\bf Rashmi R. Nayak}$^b$\footnote{e-mail: {\tt rashmi@iopb.res.in}}
{\bf \,and \,\,Kamal L. Panigrahi}$^c$\footnote{e-mail: {\tt
    Kamal.Panigrahi@roma2.infn.it}, INFN fellow}\\
\vskip .2in
{}$^a${\it Department of Physics, Stockholm University,\\
AlbaNova University Centre, SE - 106 91 Stockholm, Sweden}
\vskip .15in
{}$^b${\it Institute of Physics, Bhubaneswar 751 005, India}\\
\vskip .15in
{}$^c${\it Dipartimento di Fisica, Universita' di Roma ``Tor Vergata"\\
INFN, Sezione di Roma ``Tor Vergata", Via della Ricerca Scientifica 1\\ 
00133 Roma, Italy} 
\vspace{.7in}
\begin{abstract}
\vskip .2in
\noindent
A class of D$p$ and D$p$ -D$(p+4)$ -brane solutions are constructed in the
Penrose limit of the linear-dilaton geometry. The classical solutions
are shown to break all space-time supersymmetries. In the worldsheet
description, the branes preserve as many supersymmetries as D-branes in
flat space. It is shown that these supersymmetries do not have zero
modes on the worldsheet and hence do not admit local space-time
realizations. This indicates that, unlike the perturbative spectrum,
the D-brane spectrum of strings in the linear-dilaton pp-wave is
not similar to the flat space case.
\end{abstract}
\end{center}
\vfill

\end{titlepage}
\setcounter{footnote}{0}
\renewcommand{\thefootnote}{\arabic{footnote}}

\sect{Introduction}

String theory in plane wave backgrounds that arise as Penrose limits
of certain near-horizon geometries \cite{BFHP} is known to provide a
holographic description of certain sectors of the dual field theory
\cite{BMN}. The Penrose limits of geometries of the form $AdS_p\times
S^q$ lead to pp-wave/CFT correspondence and are of particular interest
\cite{BMN,mukhi,gomis,narain}. The introduction of D-branes in the
$AdS_p\times S^q$ and pp-wave space-times corresponds to considering
defect conformal field theories on the dual side
\cite{Bachas:2000fr,Karch:2000ct,Karch:2000gx}. For D-brane classical
solutions in such pp-wave backgrounds see also
\cite{sken}-\cite{Ohta:2003rr}. 

In this paper we are interested in a Penrose limit of the near
horizon geometry of the NS5-brane (the linear-dilaton background)
considered in \cite{go,kirit,rangamani}. String theory in this
background is related, by a coordinate transformation, to the
Nappi-Witten model \cite{nappi} which is well studied (see, for example,
\cite{Kiritsis:jk,Forgacs:1995tx,Russo:2002rq}). In particular it is
known that the worldsheet theory in the light-cone gauge can be mapped
to a theory of strings in flat background by a simple field
redefinition. However, as a result of the field redefinition, closed
strings in the redefined theory have twisted boundary conditions that
shift their oscillator frequencies. 

In \cite{rangamani} it was argued that this pp-wave limit of the
linear dilaton geometry contains the holographic description of a
high-energy and large R-charge sector of the little string theory
\cite{LST} that is dual to the linear dilaton space-time. The
similarity of string theory in the linear-dilaton pp-wave to strings
in flat background then suggested that the spectrum of little string
theory in the high-energy and large R-charge regime is very similar to
the spectrum of free strings in 10 dimensional flat space-time.

In this paper we consider D-branes embedded in the linear-dilaton
pp-wave background. The purpose is to investigate to what extent the
similarity with string theory in flat space-time persists. It is shown
that significant differences arise in the way space-time supersymmetry
is manifested. We consider a class of D$p$-branes and
(D$p$-D$p'$)-branes that contain the light-cone directions of the
pp-wave within their worldvolumes but share the transverse directions
of the pp-wave. It is shown that the classical solutions do not
preserve any space-time supersymmetry. On the other hand, in the
worldsheet description, these D-branes are shown to preserve the same
amount of supersymmetry as the corresponding D-branes in flat space.
However, the associated supercharges do not contain zero-mode pieces
and therefore do not have local realizations in space-time. One
expects this to have implications for the analogue of the defect CFT
in the context of little string theories. For related work on open
strings in NS-NS pp-wave background see \cite{Takayanagi:2002je,
Hikida:2002in,michi}.
 
The paper is organized as follows: In section 2, we present D$p$ and
(D$p$-D$p'$)-brane solutions in the background obtained as a Penrose
limit of the NS5-brane near-horizon geometry. In section 3, we show
that these solutions break all space-time supersymmetries. In section
4, we describe the D1-brane in the light-cone worldsheet theory and
show that while supersymmetry is preserved on the worldsheet it does
not admit a local space-time realization. Section 5 contains our
conclusions and discussions.

\sect{Classical solutions}

In this section, we present classical solutions corresponding to
D$p$ and D$p$-D$p'$ branes in the plane wave background with constant
NS-NS 3-form flux that arises as a Penrose limit of the near horizon
geometry of the NS5-brane (\cite{rangamani}). The string frame metric,
3-form $H$ and dilaton of the NS5-brane background are given by,
\bea
ds^2 &=& -dt^2 + dy^2_5 + f(r)(dr^2 + r^2 (d\theta^2 
+ \cos^2 \theta d\psi^2 + sin^2 \theta d\phi^2))\,, \nn\\[.1cm]
H&=&N \epsilon_3\,,\qquad e^{2\phi} = g^2_s f(r)\,,\qquad f(r) = 1+
{{Nl^2_s} \over {r^2}}\,,\nn 
\eea 
where $\epsilon_3$ is the volume form on the transverse $S_3$ and
$N$ is the NS5-brane charge. The near-horizon geometry corresponds
to the the limit $r\rightarrow 0$ which removes the $1$ in $f(r)$ 
and, on rescaling the time, leads to the linear dilaton background,
\bea
ds^2 = N l^2_s \big( -d\tilde{t}^2 + {{dr^2}\over {r^2}} +
\cos^2 \theta d\psi^2 + d\theta^2 + sin^2 \theta d\phi^2 \big)+
dy^2_5\,.\nn
\eea       
The Penrose limit is then taken with respect to the null geodesic
along the equator ($\theta=0$) of the transverse $S^3$ resulting in    
the following expressions for the metric and NS-NS 3-form
\cite{rangamani},
\be
\begin{array}{l}
ds^2 = 2dx^+dx^- -\mu^2(z_1^2+z_2^2)(dx^+)^2+\sum_{a=1}^8 dz^adz^a\,,
\\[.2cm]
H_{+12} = 2\mu\,. 
\end{array}
\label{NS5pp-bg}
\ee  
This is the background that we are interested in. It is related to the
Nappi-Witten model \cite{nappi} by a coordinate transformation and has
been recently discussed in the context of pp-waves in
\cite{go,kirit,rangamani}. In particular it was argued in
\cite{rangamani} that string theory in this background provides a
holographic description of a high-energy and high R-charge sector of
little string theory \cite{LST}. The NS-NS 2-form corresponding to the
above background will be taken as
\be
B_{+1}=\mu z^2\,,\qquad  B_{+2}=-\mu z^1\,.
\label{B}
\ee
This is related to other choices of the B-field by 2-form gauge
transformations which are symmetries of the closed string sector but
modify worldvolume gauge fields in the case of open strings.

We start by writing down the D-string solution where the brane 
worldvolume is along the light-cone directions of the plane wave, 
\bea
&& ds^2=f^{-{1\over 2}}_1[2dx^+dx^- - \mu^2\sum^2_{i=1}z^2_i(dx^+)^2]
+f^{{1\over 2}}_1\sum^8_{a=1}{(dz^a)}^2, \nn\\[.1cm]
&& H_{+12} = 2\mu \,, \qquad  F_{+-a}= \p_a f^{-1}_1\,, \nn\\[.1cm]
&& e^{2\phi} = f_1\,, \qquad f_1 = 1+{N_1 g_s\,l_s^6\over r^6}\,.
\label{D1}
\eea
For zero D-string charge this reduces to the pp-wave background
(\ref{NS5pp-bg}) and for $\mu=0$ it gives the D-string in flat space.
We have checked explicitly that the above solution satisfies type-IIB
field equations. Since the light-cone directions of the pp-wave are
within the brane worldvolume, it is possible to construct
this solution in the light-cone worldsheet theory in terms of
open-string boundary conditions. We will see below that in spite of
its simple form, this solution does not preserve any space-time
supersymmetry. The worldsheet origin of the absence of supersymmetry
will be discussed later.

Other Dp-branes for $2\le p\le 7$ can be obtained by smearing the
D-string along some of the transverse directions $z^3 \cdots z^8$ and
applying $T$-dualities. Since the D-string solution (\ref{D1}) does 
not preserve supersymmetries, smearing the background is not {\it a  
priory} justifiable (since it is not obvious that stable periodic
arrays could be built). However, one can check that the resulting 
configurations satisfy the IIA or IIB equations of motion, as the
case may be. Then, for example, the $D3$-brane solution in the NS5
pp-wave background is obtained by applying $T$-dualities along
$z^3$ and $z^4$-directions, 
\bea
&& ds^2=f^{-\frac{1}{2}}_3[2dx^+dx^- - \mu^2\sum^2_{i=1}z^2_i(dx^+)^2
+\sum^4_{\alpha=3}{(dz^\alpha)}^2] + 
f^{\frac{1}{2}}_3 \sum^8_{a=1,2;5}{(dz^a)}^2\,, \nn\\[.1cm]
&& H_{+12} = 2\mu\,, \qquad F_{+-34a}= \partial_a f^{-1}_3\,,\qquad
e^{2\phi} = 1\,, \qquad f_3 = 1+\frac{N_3 g_s\,l_s^4}{r^4}\,,
\label{D3}
\eea
with $f_3$ being the harmonic function in the transverse 6-space.

Now we present classical solutions for intersecting branes of 
type $(Dp-D(p+4))$ in the $NS5$ plane wave background. The classical 
solution for intersecting $(D1-D5)$-branes is given by,
\bea
&& ds^2={(f_1f_5)}^{-\frac{1}{2}}[2dx^+dx^- - \mu^2\sum^2_{i=1}z^2_i
(dx^+)^2] \nn\\[.1cm]  
&& \qquad \qquad \qquad \qquad \qquad 
+f_1^\frac{1}{2}f_5^{-\frac{1}{2}}\sum^6_{\alpha=3} {(dz^{\alpha})}^2 
+ {(f_1f_5)}^{\frac{1}{2}} \sum_{a=1,2,7,8}{(dz^a)}^2\,, 
\nn\\[.1cm] 
&& H_{+12} = 2\mu \,,\qquad F_{+-a} = \partial_a f^{-1}_1\,,
\qquad F_{abc}=\epsilon_{abcd}\partial_d f_5\,,
\nn\\[.1cm] 
&&e^{2\phi} = {f_1\over f_5}\,,\qquad
f_{1,5} = 1+ \frac{{N_{1,5}g_s\,l_s^2}}{r^2}\,.
\label{D1-D5}
\eea
$f_1\,,f_5$ are harmonic functions in the common transverse
directions $(z^1,z^2,z^7,z^8)$. We have checked that the above
solution satisfies all type-IIB field equations.  

Again, other solutions can be obtained from this by T-dualities, for
example, the intersecting $(D3-D7)$-branes configuration,
\bea
&& ds^2= {(f_3f_7)}^{-\frac{1}{2}}[2dx^+dx^- - \mu^2
\sum^2_{i=1}z^2_i(dx^+)^2 + \sum^4_{\alpha=3}(dz^\alpha)^2]
\nn\\[.1cm] 
&& \qquad \qquad\qquad \qquad\qquad \qquad
+{f_3}^{\frac{1}{2}}{f_7}^{-\frac{1}{2}}\sum^8_{\beta=5}{(dz^{\beta})}^2
+ {(f_3f_7)}^{\frac{1}{2}} \sum^2_{i=1}{(dz^i)}^2 \,, \nn\\[.1cm]
&& H_{+12} = 2\mu\,,\qquad F_{+-34i} =\partial_i f^{-1}_3\,, 
    \qquad \p_i \chi = \epsilon_{ij}\p_j f_7  \nn\\[.1cm]
&& e^{2\phi} = f^{-2}_7\,, \qquad  f_{3,7} = 1+ c\ln {r\over l}\,,
\label{D3-D7}
\eea
More intersecting brane solutions can be obtained by using 
$T$-dualities along $z^3,\cdots,z^8$ directions in (\ref{D1-D5}).
D-branes intersecting at angle can also be obtained following
\cite{rashmi,myers1}, however, we will skip these details.

\sect{Supersymmetry analysis of classical solutions}

Let us now look at the supersymmetry of the D-brane solutions
in the NS5 near-horizon plane wave by solving the type IIB Killing
spinor equations. The supersymmetry variations of dilatino
and gravitino fields of type IIB supergravity in the string frame are
given by \cite{schwarz,fawad}, 
\begin{eqnarray}
\delta \lambda_{\pm} &=& {1\over2}(\Gamma^{\mu}\partial_{\mu}\phi \mp
{1\over 12} \Gamma^{\mu \nu \rho}H_{\mu \nu \rho})\epsilon_{\pm} + {1\over
  2}e^{\phi}(\pm \Gamma^{\mu}F^{(1)}_{\mu} + {1\over 12} \Gamma^{\mu \nu
  \rho}F^{(3)}_{\mu \nu \rho})\epsilon_{\mp}\,,\\
\label{dilatino}
\delta {\Psi^{\pm}_{\mu}} &=& \Big[\partial_{\mu} + {1\over 4}(w_{\mu
  \hat A \hat B} \mp {1\over 2} H_{\mu \hat{A}
  \hat{B}})\Gamma^{\hat{A}\hat{B}}\Big]\epsilon_{\pm} \cr
& \cr
&+& {1\over 8}e^{\phi}\Big[\mp \Gamma^{\lambda}F^{(1)}_{\lambda} - {1\over 3!}
\Gamma^{\lambda \nu \rho}F^{(3)}_{\lambda \nu \rho} \mp {1\over 2.5!}
\Gamma^{\lambda \nu \rho \alpha \beta}F^{(5)}_{\lambda \nu \rho \alpha
  \beta}\Big]\Gamma_{\mu}\epsilon_{\mp}\,,
\label{gravitino}
\end{eqnarray}
where $\mu, \nu ,\rho, \lambda$ are ten dimensional space-time
indices, and hated indices refer to the Lorentz frame.

Note that for the background (\ref{NS5pp-bg}), the vanishing of the
above supersymmetry variations leads to the Killing spinors,
\be
\epsilon^{(bg)}_\pm = e^{\pm\frac{\mu}{2}x^+\Gamma^{\hat 1\hat 2}}
\epsilon^{(0)}_\pm \,,\qquad \Gamma^{\hat +} \epsilon^{(0)}_\pm=0\,, 
\label{ks-bg}
\ee
where $\epsilon^{(0)}_\pm$ are constant spinors.

As for the D-brane solutions (\ref{D1})-(\ref{D3-D7}) in the
background (\ref{NS5pp-bg}), it suffices to analyze the basic D-string
solution (\ref{D1}). The vanishing of the dilatino variations give
\bea
\frac{f_{1,\hat a}}{f_1}\, \Gamma^{\hat a}\Big(\epsilon_{\pm} - 
\Gamma^{\hat +\hat -}\epsilon_{\mp}\Big) \mp
2\mu~f^{-{1/4}}_1 \Gamma^{\hat +\hat 1\hat 2}\epsilon_{\pm} = 0\,,
\label{dil}
\eea
where hats denote Lorentz frame indices and $\hat a=1\cdots 8$.  On
multiplying by $\Gamma^{\hat +}$ these reduce to $\Gamma^{\hat +}
(\epsilon_+ +\epsilon_-)=0$. Also, adding the upper and lower sign
equations in (\ref{dil}) leads to $\Gamma^{\hat +}(\epsilon_+
-\epsilon_-)=0$, while subtracting the two gives
$\Gamma^{\hat-}(\epsilon_+-\epsilon_-)= 0$. The only non-trivial
solution of these equations is  
\bea
\epsilon_+ =\epsilon_- =\epsilon \,,\qquad 
\Gamma^{\hat +}\epsilon =0\,.
\label{eps}
\eea
The second condition is a common feature of pp-wave backgrounds, while
the first is enforced by the presence of the D-string.
The only component of the gravitini variations consistent with the
above is \footnote{In our conventions, $\Gamma^{\hat +\hat -}
\Gamma^{\hat +}= -\Gamma^{\hat +}\Gamma^{\hat +\hat -}=
\Gamma^{\hat +}$ and $\Gamma^{\hat +\hat -}\Gamma^{\hat -}= 
-\Gamma^{\hat -}\Gamma^{\hat +\hat -}=-\Gamma^{\hat -}$.}  
\bea
\delta \psi_-^{\pm} \equiv \partial_{-}\epsilon_{\pm} + {1\over 8}
{f_{1,\hat a}\over {f^{5/4}_1}}\,\Gamma^{\hat a}\Gamma^{\hat +}
\epsilon_{\mp} = 0,
\eea
giving $\partial_-\epsilon =0$. Among the remaining gravitini
variations let us first consider 
\bea
\delta \psi_+^{\pm} &\equiv& \partial_{+}\epsilon_{\pm} 
+{1\over 8} f_1^{-5/4}f_{1,\hat a}\, \Gamma^{\hat a}
\left(\Gamma^{\hat -}+\frac{1}{2}\mu^2 z_iz_i \Gamma^{\hat +}\right)
(\epsilon_\pm - \epsilon_\mp) \nn\\[.1cm]
&&\qquad\qquad 
+\,{1\over 2}\mu^2 z_{\hat i} f_1^{-1/4}\Gamma^{\hat i}\,
\Gamma^{\hat +}\epsilon_{\pm} \mp {\mu \over 2} f^{-1/2}_1 
\Gamma^{\hat 1\hat 2} \epsilon_{\pm} =0 \,.
\label{g1}
\eea
On imposing (\ref{eps}), these reduce to the two equations
$\partial_{+}\epsilon \mp {\mu \over 2} f^{-1/2}_1 \Gamma^{\hat 1\hat
2} \epsilon =0$ which are clearly inconsistent which each other. Even
if one of the above two equations is relaxed, the solution
$\epsilon(x^+, x^a)$ is not consistent with the vanishing of the
remaining gravitini variations,
\be
\delta \psi_c^{\pm} \equiv \partial_{c}\epsilon_{\pm} 
- {1\over 8}\delta_{c\hat b}{f_{1,\hat a}\over{f^{3/4}_1}} 
\left(\Gamma^{\hat a\hat b}\epsilon_{\pm} + \Gamma^{\hat a}\Gamma^{\hat b} 
\Gamma^{\hat +\hat -}\epsilon_{\mp} \right)
\mp {\mu\over 2}\delta_{c\hat i}\Gamma^{\hat +\hat i}\epsilon_{\pm}
= 0 \,,
\label{g2}
\ee
due to the specific form of its $x^+$-dependence. This shows that the
D-string solution in equation (\ref{D1}) does not preserve any of the
space-time supersymmetries. Similarly one can show that, for the same
reasons as above, the intersecting D-brane solutions in the NS5
pp-wave background presented in the previous section also do not
preserve any space-time supersymmetry. This may sound a little
puzzling since the worldsheet theory in the background
(\ref{NS5pp-bg}) is closely related to the theory in flat space-time.
Below we will see that the small difference between the two is enough
to destroy all space-time supersymmetries in the D-brane sector.

\sect{Worldsheet analysis of D-brane supersymmetry}

String propagation in the NS5 pp-wave background is related to the
Nappi-Witten model \cite{nappi} and has been extensively studied
\cite{kirit,Kiritsis:jk,Forgacs:1995tx,Russo:2002rq,
Takayanagi:2002je,Hikida:2002in,michi}. The worldsheet theory is
greatly simplified in the light-cone Green-Schwarz description where
by a field redefinition the action reduces to that of strings in flat
space-time. The similarity of closed string spectrum in this pp-wave
background to that in flat space-time has been emphasized in the
context of the dual little string theory \cite{rangamani}. However, as
we have seen above, at the level of classical solutions, the
supersymmetry of D-brane excitations is very different from flat
background. We now discuss this issue for the D-string solution from
the worldsheet point of view. The other cases are similar.

The light-cone gauge Green-Schwarz action for strings in the linear
dilaton background (\ref{NS5pp-bg}) can be worked out using the IIB GS
action in \cite{Cvetic:1999zs} or can be directly read off from
\cite{Russo:2002rq}, as\footnote{For the $B$-field we use the form 
(\ref{B}) which, for closed strings, is equivalent to other choices
related to it by 2-form gauge transformations, {\it e.g.}, $B_{12}=2\mu
x^+$. For open strings there is some ambiguity which we fix by using
(\ref{B}) and setting the worldvolume gauge fieldstrength to zero.
This is consistent with our classical solution and allows fixing the
light-cone gauge on the worldsheet.} 
\bea   
{\cal S} &=& \int d^2\sigma \Big[\sum_{i=1}^2(\p_+ X^i\p_- X^i - 
m^2 X^iX^i) -2m\,(X^1\p_\sigma X^2 -X^2\p_\sigma X^1) \nonumber \\ 
&& \qquad +\sum_{a=3}^8 \p_+ X^a\p_- X^a \nonumber \\ 
&& \qquad +i\,(S^T_L\,\p_- S_L + \frac{m}{2}\, S^T_L\,\gamma^{12} S_L)
 +i\, (S^T_R\,\p_+ S_R - \frac{m}{2}\, S^T_R\,\gamma^{12} S_R)\Big]\,.
\label{Slc}
\eea
Here $S_{L,R}$ are $SO(8)$ spinors in, say, ${\bf 8}_s$, $m=\mu p^+$
and $\p_\pm = \p_\tau\pm\p_\sigma$. For closed strings the boundary
conditions are periodicity in the $X$'s and $S_{L,R}$. For open
strings, the boundary terms in the variation of the action can be set
to zero by the D1-brane boundary conditions
\be    
\delta X^i\vert_{\sigma =0,\pi}=0\,,\quad 
\delta X^a\vert_{\sigma =0,\pi}=0\,,\quad 
S_L - \Omega S_R\vert_{\sigma =0,\pi}\,,\quad (\Omega^T\Omega=1)\,, 
\label{D1bc}
\ee
which are imposed for all $\tau$. $\Omega$ is a constant matrix to be
determined by the requirement that supersymmetry transformations keep
the boundary conditions invariant.  

The action (\ref{Slc}) can be recast as a flat-space action by field
redefinitions. In the bosonic sector the required transformation is 
a rotation by an angle $m\sigma$ \cite{Forgacs:1995tx},
\be
Y^1+iY^2 = e^{im\sigma} (X^1+iX^2)\,.
\label{Xredef}
\ee 
As for the fermions, there are two possible choices. For closed
strings, where boundary conditions do not relate $S_L$ and $S_R$, 
it is convenient to use 
\be 
S_L=\exp(-\frac{m}{2}\tau\gamma^{12})\hat S_L^{closed}\,,\qquad
S_R=\exp(\frac{m}{2}\tau\gamma^{12})\hat S_R^{closed}\,.
\ee
Then, $\hat S_L^{closed}(\tau+\sigma)$ and $\hat S_R^{closed}
(\tau-\sigma)$ satisfy the free field equations with periodic boundary
conditions along $\sigma$. In the light-cone gauge, the zero mode part
of these can be written entirely in terms of the space-time variable  
$X^+=p^+\tau$ which is consistent with the structure of Killing
spinors of the NS5 pp-wave background (\ref{ks-bg}).     

However for open strings where (\ref{D1bc}) should hold for all
$\tau$, we find it convenient to define
\be 
\hat S_{L,R}=\exp(-\frac{m}{2}\sigma\gamma^{12}) S_{L,R}\,,  
\label{Sredef}
\ee
which is a spinor representation of (\ref{Xredef}). The action then
takes the form 
\be   
{\cal S}= \int d^2\sigma \Big[\sum_{i=1}^2 \p_+ Y^i\p_- Y^i  
+\sum_{a=3}^8 \p_+ X^a\p_- X^a 
+i\,(\hat S^T_L\,\p_-\hat S_L +\hat S^T_R\,\p_+\hat S_R )\Big]\,.
\label{Sflat}
\ee
The equations of motion are solved by $\hat S_L(\tau+\sigma)$,
$\hat S_R(\tau-\sigma)$ and $Y^i=Y^i_L(\tau+\sigma)+
Y^i_R(\tau-\sigma)$ subject to boundary conditions that follow
from (\ref{D1bc}).  

This action is invariant under the full set of supersymmetry
transformations generated by spinors $\hat\eta_L(\tau+\sigma)$,
$\hat\eta_R(\tau-\sigma)$ in the ${\bf 8}_s$ and the spinors
$\hat\epsilon_{L,R}$ in the ${\bf 8}_c$ of $SO(8)$ subject to boundary
conditions (see for example \cite{Green:sp}). Explicitly, one has the
flat background kinematic supersymmetries, 
\be
\delta_{\eta} \hat S_{L} = \hat\eta_{L}\,,\quad
\delta_{\eta} \hat S_{R} = \hat\eta_{R}\,,    
\ee
and the dynamical supersymmetries ($i=1,2; a=3,\cdots 8$),
\bea
&\delta_{\epsilon}\hat S_R=-(\p_-Y^i\gamma_i+\p_-X^a\gamma_a)\hat\epsilon_R
\,,\qquad 
\delta_{\epsilon}\hat S_L=(\p_+Y^i\gamma_i+\p_+X^a\gamma_a)\hat\epsilon_L\,,&
\\[.2cm]
&\,\,\,\delta_{\epsilon}Y^i=2i(\hat\epsilon^{T}_R\gamma^i\hat S_R-
\hat\epsilon^T_L  \gamma^i\hat S_L)\,,\,\, \qquad \quad
\delta_{\epsilon}X^a=2i(\hat\epsilon^T_R\gamma^a\hat S_R-\hat\epsilon^T_L 
\gamma^a\hat S_L)\,.&
\label{susyff}
\eea
The boundary conditions (\ref{D1bc}) along with the invariance of the
action (\ref{Sflat}) under the above supersymmetry transformations lead
to the following boundary conditions on the redefined flat-space
quantities, 
\be    
\p_\tau Y^i\vert_{\sigma =0,\pi}=0\,,\quad 
\hat S_L-\hat\Omega\hat S_R\vert_{\sigma =0,\pi}\,,\quad 
\hat \eta_L-\hat\Omega\hat \eta_R\vert_{\sigma =0,\pi}\,,\quad 
\hat \epsilon_L-\hat\Omega\hat \epsilon_R\vert_{\sigma =0,\pi}\,, 
\label{D1bcff}
\ee
where $\hat\Omega=\exp(-m\sigma\gamma^{12}/2) \Omega
\exp(m\sigma\gamma^{12}/2)$. It is easy to verify that the
supersymmetry transformations (\ref{susyff}) keep the boundary
conditions invariant provided $\Omega=\hat\Omega =1$. 

Thus from the worldsheet point of view, the D-string in the NS5
pp-wave background has as many space-time supersymmetries as the
D-string in flat space. However, when written in terms of the
variables $X^i$ and $S_{L,R}$, it becomes evident that all these
supersymmetry transformations acquire a dependence on the worldsheet
coordinate $\sigma$ in such a way that they do not contain
$\sigma$-independent zero-mode pieces. Explicitly, the supersymmetry
transformations in terms of the original variables become,
\bea
\delta_{\eta}S_{L}&=&\eta_{L}\,, \\
\delta_{\eta}S_{R}&=& \eta_{R}\,,\\    
\delta_{\epsilon}S_R&=&-(\p_-X^i\gamma^i+\p_-X^a\gamma_a -
m\,\epsilon_{ij}X^i\gamma^j)\, \epsilon_R\,, \\
\delta_{\epsilon}S_L&=&(\p_+X^i\gamma_i+\p_+X^a\gamma_a + 
m\,\epsilon_{ij}X_i\gamma^j)\,\epsilon_L\,,\\
\delta_{\epsilon}X^i&=&2i(\epsilon^{T}_R\gamma^i S_R-\epsilon^T_L 
\gamma^i S_L)\,,
\eea  
which are compatible with the boundary conditions (\ref{D1bc}). The 
transformation parameters $\eta$ and $\epsilon$ are related to
$\hat\eta$ and $\hat\epsilon$ by equations similar to (\ref{Sredef}). 
That these do not contain $\sigma$-independent pieces (zero-modes
along the string) is easy to see. For example, before imposing the
boundary condition, $\eta_{L,R}$ have zero modes, 
\bea 
(\eta_L)_0=\frac{1}{2}(1-i\gamma^{12})\hat\eta_{L,-\frac{m}{2}}e^{-im\tau/2}
+\frac{1}{2}(1+i\gamma^{12})\hat\eta_{L,\frac{m}{2}}e^{im\tau/2}\,,
\label{zeromodeL}\\ 
(\eta_R)_0=\frac{1}{2}(1-i\gamma^{12})\hat\eta_{R,\frac{m}{2}}e^{im\tau/2}+
\frac{1}{2}(1+i\gamma^{12})\hat\eta_{R,-\frac{m}{2}}e^{-im\tau/2}\,,
\label{zeromodeR}
\eea
where $\hat\eta_n$ are the oscillators in the expansion of $\hat\eta$.
But for the boundary condition $\eta_L-\eta_R\vert_{\sigma=0,\pi}=0$
to hold at all $\tau$, one must have,
\bea
(1+i\gamma^{12})\hat\eta_{L,\frac{m}{2}}  &=&
(1-i\gamma^{12})\hat\eta_{R,\frac{m}{2}} \,, \nonumber\\
(1-i\gamma^{12})\hat\eta_{L,-\frac{m}{2}}   &=& 
(1+i\gamma^{12})\hat\eta_{R,-\frac{m}{2}} \,,\nonumber
\eea
It is easy to see that the solutions to these equations, when
substituted back in (\ref{zeromodeL}) and (\ref{zeromodeR}), lead to 
\be 
(\eta_L)_0= 0 \,,\qquad
(\eta_R)_0= 0 \,.
\ee
Thus the supersymmetry transformations that are consistent with the
boundary conditions do not have zero modes. In other words, this means
that the supercharges do not contain $\sigma$-independent components
and therefore cannot be expressed in terms of space-time variables
alone. Consequently the worldsheet supersymmetries of D1-brane in NS5
pp-wave background are not manifest in the space-time description,
consistent with the space-time supersymmetry analysis. Similar
situations were encountered in \cite{sken} in more involved setups.
The example considered here is the simplest and most drastic
realization of the phenomenon.

In fact, situations where worldsheet symmetries do not lead to
space-time symmetries due the $\sigma$-dependence of charges (or
equivalently, absence of zero modes in the transformation) have been
known for a long time, mostly in relation to T-duality. An example
where symmetries and supersymmetries at the current algebra level do
not have a space-time realization was discussed in
\cite{Antoniadis:1994sr} and its relation to T-duality was clarified
in \cite{Bakas:1995hc}. The more general situation where supersymmetry
transformations could acquire a $\sigma$-dependence as a result of
T-duality was analyzed in \cite{Hassan:1995je}. In all these cases
the symmetry exists at the worldsheet level but is not locally
realized on the space-time fields.

\sect{Summary and Discussion}

We have constructed Dp-brane solutions in the background of a pp-wave 
limit of the linear dilaton geometry. The configurations are such that   
the light-cone directions of the pp-wave fall within the worldvolume
of the D-brane and the directions transverse to the pp-wave are also
transverse to the brane. It is shown that these classical solutions do
not preserve any space-time supersymmetry. 

Such branes can also be constructed in the worldsheet theory in the
pp-wave background. In the light-cone gauge, this can be mapped to a
theory of strings in flat space-time by simple field redefinitions. It
is shown that in the worldsheet description, the D-branes preserve as
much supersymmetry as the corresponding D-branes in flat space-time.
The contradiction between the space-time and worldsheet results is
resolved by showing that all Fourier modes of the allowed
supersymmetry parameters depend on the worldsheet coordinate $\sigma$,
while a local description in terms of space-time fields should be
blind to the extension of the string. As a result it is not possible
to express any mode of the supersymmetry parameters in terms of the
space-time variables alone, hence the absence of ordinary space-time
supersymmetry. This is similar to certain situations discussed in
\cite{sken} and in fact provides the simplest and most drastic example
of this effect. More generally, this is an example of a phenomenon
encountered earlier in the context of T-duality
\cite{Antoniadis:1994sr,Bakas:1995hc,Hassan:1995je}.

String theory in the linear dilaton pp-wave background is argued to
provide a holographic description of a certain high-energy and large
R-charge sector of the associated little string theory
\cite{rangamani}. The spectrum in this sector was argued to be similar
to the spectrum of closed strings in flat 10-dimensional space based
on the similarity between the latter and closed strings on the linear
dilaton background. However, our discussion shows that this similarity
does not extend to string spectrum in the presence of D-branes in a
straightforward way. It is interesting to investigate the consequence
of this for the defect CFT analogue the little string theory.

Our results can be compared with earlier work on D-branes in the
pp-wave limit of $AdS_3\times S^3$ geometry with NS-NS 3-form
background \cite{kamal, michi}. In this case the structure of the
NS-NS 3-form is such that the classical solutions preserve some
supersymmetries. An inspection of the worldsheet theory along the
lines discussed in section 4 shows that the same feature of the NS-NS
3-form also insures that some of the worldsheet fermions have zero
modes resulting in the observed space-time supersymmetries. The fate
of the remaining supersymmetries is as discussed above.

\vskip .5cm
\noindent
{\large\bf{Acknowledgment:}} 

We would like to thank B. Eden, Ya. Stanev and especially M. Bianchi
and A. Sagnotti for various useful discussions. The research of K.P.
was supported in part by I.N.F.N., by the E.C. RTN programs
HPRN-CT-2000-00122 and HPRN-CT-2000-00148, by the INTAS contract
99-1-590, by the MURST-COFIN contract 2001-025492 and by the NATO
contract PST.CLG.978785.


\end{document}